\documentstyle[12pt]{article}

\begin{document}
\title{The Higgs boson mass bound in MSSM broken at high scale}
\author{N.V.Krasnikov\\ INR RAS, Moscow 117312 \\
G.Kreyerhoff and R.Rodenberg\\
RWTH Aachen, III. Physikalisches Institut \\ Abteilung f\"ur 
Theoretische Elementarteilchenphysik\\ Physikzentrum\\52056 Aachen\\Germany}

\date{PITHA 97/25}
\maketitle
\begin{abstract}
We study the dependence of the Higgs boson mass on the supersymmetry 
breaking scale in the minimal supersymmetric extension of the 
standard Weinberg-Salam model. In particular, we find that for supersymmetry 
breaking scale $10^8$ GeV $\leq M_s \leq 10^{16}$ GeV  and for 
$m^{pole}_{top} = 175 \pm 5$ GeV the Higgs boson mass is 120 GeV $ \leq 
m_h \leq 160 $ GeV.

\end{abstract}

\newpage
At present one of the most urgent problems in high energy physics is the 
search for the Higgs boson. 
The lower LEP1 bound on the Higgs boson mass is \cite{1}
\begin{equation}
m_h > 66\,GeV \,.
\end{equation}
In standard  Weinberg-Salam model there are several theoretical
bounds on the Higgs boson mass :

(i) Tree level unitarity requirement leads to $m_h \leq 1$ TeV
\cite{2}.

(ii) The requirement of the absence of the Landau pole singularity
for the effective Higgs self-coupling constant for energies up to
$10^{14}$  GeV gives $m_h \leq 200$ GeV for $m_{top}^{pole} \leq 200$ GeV
\cite{3}.

(iii) The vacuum stability requirement leads to the lower
bound on the Higgs boson mass which depends on the top quark
mass \cite{4}.

The minimal supersymmetric extension of the standard model (MSSM) 
\cite{5} predicts at tree level that the lightest Higgs boson has to be 
lighter than the Z-boson \cite{5}. Radiative corrections slightly 
increase the value of the lightest Higgs boson mass \cite{6}.

In our previous paper \cite{7} we studied the dependence of the Higgs boson 
mass on the scale  of supersymmetry breaking in MSSM using one loop 
renormalization group equations for the effective coupling constants. 
In this note we reanalize this problem using two loop renormalization 
group equations for the effective coupling constant and recent more 
accurate world average value of the strong coupling constant 
$\alpha_{s}(M_z) =0.118 \pm 0.003$  \cite{8} and the value of the 
top quark mass $m_{top}^{pole} = 175 \pm 6$ Gev \cite{9}.

Our main assumption is that the standard Weinberg-Salam model originates 
from its minimal supersymmetric extension which is explicitly broken due to 
soft supersymmetry breaking terms at scale $M_s$. The tree level Higgs 
potential in the MSSM model with general soft supersymmetry breaking 
terms is given by \cite{5}
\begin{equation}
V = V_0 + V_{soft},
\end{equation}
\begin{equation}
V_0 = (g^2_1 + g^2_2)(H^{+}_{1}H_{1} - H^{+}_{2}H_{2})^2/8 + 
g^2_2|H^{+}_{1}H_{2}|^{2}/2,
\end{equation}
\begin{equation}
V_{soft} = m^2_1 H_1^{+}H_1 + m^2_2H^{+}_{2}H_{2} + 
m^2_{3}(H^{T}_1i \tau_{2}H_2 
+ h.c.) .
\end{equation}
Here $g_1$ and $g_2$ are the $U(1)$ and $SU(2)$ gauge coupling constants 
and the Higgs doublets $H_1$ and $H_2$ couple with $q = -1/3$ and $q = 2/3$ 
quarks respectively. We assume that one of the combinations of the $H_1$ 
and $H_2$ 
\begin{equation}
H_{light} = H_2 \cos(\phi) + i\tau_{2}H^{+}_{1}\sin(\phi) 
\end{equation}
is relatively light, $m_{light} \approx O(M_z)$, whereas the other 
orthogonal combination 
\begin{equation}
H_{heavy} = -H_2 \sin(\phi) + i \tau_{2}H^{+}_{1}\cos(\phi)
\end{equation}
acquires a mass $m_{heavy} \approx O(M_s)$. We also assume that the 
masses of the superpartners of ordinary particles are of the order of 
$O(M_s)$. It is clear that for such scenario for $M_s \geq O(1)$ TeV it 
is necessary to have fine tuning among the soft supersymmetry breaking 
terms. 

At scales lower than the supersymmetry breaking scale 
$M_s$ we have the standard Weinberg-Salam model with the single Higgs 
isodoublet $H = H_{light}$. The crusial point is that the self-interaction 
effective coupling constant $\bar{\lambda}(M_s)$ at scale $M_s$ is
\begin{equation}
0 \leq \bar{\lambda}(M_s) = (\bar{g}^2_1(M_s) + \bar{g}^2_2(M_s))
(\cos(2\phi))^2/4 \leq (\bar{g}^2_1(M_s) + \bar{g}^2_2(M_s))/4
\end{equation}
To preserve the supersymmetry the gauge couplings $\bar{g}_{1}(M_s)$ and 
$\bar{g}_{2}(M_s)$ have to be calculated within the $\bar{DR}$-scheme 
\cite{10}. The relation between the gauge coupling constants in the 
$\bar{MS}$-scheme and $\bar{DR}$-scheme has the form \cite{10}
\begin{equation}
\frac{1}{\alpha_{i_{\bar{MS}}}} = \frac{1}{\alpha_{i_{\bar{DR}}}} 
+  \frac{C_2(G)}{12\pi},
\end{equation}
where $C_2(G)$ is the quadratic Casimir operator for the adjoint 
representation ($C_2(SU(N)) = N$. 

So the assumption that standard Weinberg-Salam model originates 
from its supersymmetric extension with the supersymmetry broken at 
scale $M_s$ allows us to obtain non-trivial information about the 
low energy effective Higgs self-coupling constant and hence to obtain 
nontrivial information about the Higgs boson mass. To relate the high 
energy value of the Higgs self-interaction effective coupling constant 
$\bar{\lambda}(M_s)$ with the low energy value of $\bar{\lambda}(M_t)$ 
it is necessary to use the renormalization group requations.       
The renormalization group equations for the effective coupling
constants in neglection of all Yukawa coupling constants except
top-quark Yukawa coupling constant
in one-loop approximation read
\begin{equation}
\frac{d\bar{g}_3}{dt} = -7\bar{g}^3_3\,,
\end{equation}
\begin{equation}
\frac{d\bar{g}_2}{dt} = -(\frac{19}{6})\bar{g}^3_2\,,
\end{equation}
\begin{equation}
\frac{d\bar{g}_1}{dt} = (\frac{41}{6})\bar{g}^3_1\,,
\end{equation}
\begin{equation}
\frac{d\bar{h}_t}{dt} = (\frac{9\bar{h}^2_t}{2} -8\bar{g}^2_3 -
\frac{9\bar{g}^2_2}{4} -\frac{17\bar{g}^2_1}{12})\bar{h}_t\,,
\end{equation}
\begin{equation}
\frac{d\bar{\lambda}}{dt} = 12(\bar{\lambda}^2 + (\bar{h}^2_t -
\frac{\bar{g}2_1}{4} - \frac{3\bar{g}^2_2}{4})\lambda -\bar{h}^4_t
+ \frac{\bar{g}^4_1}{16} + \frac{\bar{g}^2_1\bar{g}^2_2}{8} +
\frac{3\bar{g}^4_2}{16}\,,
\end{equation}
\begin{equation}
t = (\frac{1}{16\pi^2})\ln{(\mu/m_Z)}\,.
\end{equation}
Here $\bar{g}_3$, $\bar{g}_2$ and $\bar{g}_1$ are the $SU(3)$,
$SU(2)$ and $U(1)$ gauge coupling, respectively, and $\bar{h}_t$ is
the top quark Yukawa coupling constant. In our numerical analysis we 
studied the renormalization group equations for the effective 
coupling constants in two loop approximation \cite{11} in 
$\bar{MS}$-scheme.  We used the following central values for the initial 
effective coupling constants at $M_z$-scale \cite{8},\cite{12},\cite{13}:
\begin{equation}
\alpha_3(M_z)_{\bar{MS}} = 0.118 \pm 0.003,
\end{equation}
\begin{equation}
\sin^{2}(\theta_{W})(M_z) = 0.2320 \pm 0.0005,
\end{equation}
\begin{equation}
(\alpha_{em,{\bar{MS}}}(M_z))^{-1} = 127.79  \pm 0.13
\end{equation}
For boundary condition (7) for the Higgs self-coupling constant 
$\bar{\lambda}(M_s)$ we have integrated numerically the renormalization group 
equations in two loop approximation. Also we took into account one loop 
correction to the Higgs boson mass \cite{14} (running Higgs boson mass 
$\bar{m}_h(\mu) = \sqrt{\bar{\lambda}(\mu)v}$ does not coincide with 
pole Higgs boson mass. We used two loop formulae of ref.{15} 
which relate the running top quark mass with pole top quark mass. Our 
results for the Higgs boson mass for different values of $M_s$ and 
$m^{pole}_{top}$ are presented in Table 1. Here $k =0$ corresponds to 
the boundary condition $\bar{\lambda}(M_s) = 0$ and $k=1$ corresponds 
to the boundary condition $\bar{\lambda}(M_s) = \frac{1}{4}
(\bar{g}^{2}_{1}(M_s) + \bar{g}^{2}_{2}(M_s))$. We have found that 
our numerical results practically do not depend on the uncertainties in the 
determination of the electroweak couplings at $M_z$-scale and also on the 
use of the boundary condition (6) for electroweak coupling constants 
in the $\bar{DR}$-scheme instead of the $\bar{MS}$-scheme. The uncertainty 
in the determination of the strong coupling constant at 
$M_z$-scale leads to the 
uncertainty in the determination of the Higgs boson mass less than 2 GeV. 
The dependence of the Higgs boson mass on the scale of supersymmetry 
breaking $M_s$ is very weak in the interval $10^{8}$ GeV $ \leq M_s 
\leq 10^{16}$ GeV. 

From the requirement of the absence of Landau pole singularity for 
the Higgs self-coupling constant 
$\bar{\lambda}(\Lambda)$ for the scales  up to $\Lambda = (10^{3}; 10^{4};
10^{6}; 10^{8}; 10^{10}; 10^{12}; 10^{14})$ GeV
(to be precise we require that at the scale $\Lambda$ the Higgs
self-coupling constant is $\frac{\bar{\lambda}^{2}(\Lambda)}{4\pi} \leq
1$) we have found the upper bound on the Higgs boson mass $m_h \leq
400; 300; 240; 200; 180; 170; 160)$ GeV, respectively. 

It should be noted that
in nonminimal supersymmetric electroweak models, say in the model
with additional gauge  singlet $\sigma$ we have due to the term
$k\sigma H_1 i \tau_{2}H_2$  in the superpotential an additional
term $k^2|H_1i\tau_{2}H_{2}|^2$ in the potential and as a consequence
our boundary condition for the Higgs self-coupling constant has to be
modified, namely
\begin{equation}
\bar{\lambda}(M_s) = \frac{1}{4}(\bar{g}^2_1(M_s) + \bar{g}^2_2(M_s))
\cos^2(2\varphi)
+ \frac{1}{2}\bar{k}^2(M_s)\sin^{2}(2\varphi) \geq 0
\end{equation}
The boundary condition  (18) depends on unknown coupling constant
$\bar{k}^2(M_s)$. However, it is very important to stress that for all
nonminimal supersymmetric models broken to standard Weinberg-Salam
model at scale $M_s$ the effective Higgs self-coupling constant
$\bar{\lambda}(M_s)$  is  non-negative that is a direct
consequence of the non-negativity of the effective potential
in supersymmetric models. Therefore, the vacuum stability
requirement results naturally if supersymmetry is broken at
some high scale $M_s$ and at lower scales standard Weinberg-Salam
model is an effective theory. 

As it follows from our results for 170 GeV $\leq m^{pole}_{top} \leq $
180 GeV and for $M_s \leq $ 1 TeV the Higgs boson mass is less than 120 GeV, 
while for the same values of the top quark mass and for the supersymmetry 
breaking scale $M_s \geq 10^{8}$ GeV the Higgs boson mass is larger than 
120 GeV. Therefore, for such values of the top quark mass the measurement 
of the Higgs boson mass will discriminate standard scenario with low energy 
broken suppersymmetry and scenario with standard Weinberg-Salam model 
valid up to very high scale \cite{16}.  Moreover, 
for such values of the top quark mass the discovery at LEP2 
the Higgs boson with the mass lighter than 85 GeV would mean that the 
scale of new physics is less than 5 TeV. 

We are indebted to RFFI-DFG research program project No. 436 RUS 
113/227/0 which made possible our collaboration. 

\newpage

Table 1. The dependence of the Higgs boson mass on the values of
$M_s$, $m_{top}^{pole} \equiv m_t$ and $k =0,1$. Everything except k is in GeV.

\begin{center}
\begin{tabular}{|l| |l| |l| |l| |l| |l| |l| |l| |l| |l| |l|}
\hline
$m_t$ & $165$ & $165$ & $170$ & $170$ & $175$ & $175$ & $180 $ & $180$ & $185$ &
$185$ \\
\hline
$M_S$  &k=0 &k=1& k=0 & k=1 &k=0& k=1 &k=0& k=1& k=0& k=1\\
\hline
$10^{3}$&69&111&74&114&78&117&83&120&88&123\\
\hline
$10^{3.5}$&81&117&86&120&92&124&98&128&104&132\\
\hline
$10^{4}$&89&121&95&125&101&130&108&134&114&139\\
\hline
$10^{6}$&105&129&113&135&121&141&129&147&137&153\\
\hline
$10^{8}$&112&132&120&138&129&147&138&152&146&159\\
\hline
$10^{10}$&115&133&124&140&133&147&142&154&151&161\\
\hline
$10^{12}$&117&134&126&141&136&147&145&154&154&161\\
\hline
$10^{14}$&118&134&127&141&132&148&147&156&156&164\\
\hline
$10^{16}$&118&134&128&141&138&148&148&156&158&164\\
\hline
\end{tabular}
\end{center}

\end{document}